\documentstyle[12pt]{article}
\advance \topmargin by -\headheight
\advance \topmargin by -\headsep

\evensidemargin \oddsidemargin
\begin{document}

\topmargin -.6in

%%      MACROS.TEX
%   Last revision , June 9, 1992
%
%
%               macros formatting and equations
\def\rf#1{(\ref{eq:#1})}
\def\lab#1{\label{eq:#1}}
\def\nonu{\nonumber}
\def\br{\begin{eqnarray}}
\def\er{\end{eqnarray}}
\def\be{\begin{equation}}
\def\ee{\end{equation}}
\def\foot#1{\footnotemark\footnotetext{#1}}
\def\lb{\lbrack}
\def\rb{\rbrack}
\def\llangle{\left\langle}
\def\rrangle{\right\rangle}
\def\blangle{\Bigl\langle}
\def\brangle{\Bigr\rangle}
\def\llbrack{\left\lbrack}
\def\rrbrack{\right\rbrack}
\def\lcurl{\left\{}
\def\rcurl{\right\}}
\def\({\left(}
\def\){\right)}
\def\v{\vert}
\def\bv{\bigm\vert}
\def\Bgv{\;\Bigg\vert}
\def\bgv{\bigg\vert}
\def\lskip{\vskip\baselineskip\vskip-\parskip\noindent}
\relax

%                     common physics symbols
\def\tr{\mathop{\rm tr}}
\def\Tr{\mathop{\rm Tr}}
\def\partder#1#2{{{\partial #1}\over{\partial #2}}}
\def\funcder#1#2{{{\delta #1}\over{\delta #2}}}
\def\me#1#2{\left\langle #1\right|\left. #2 \right\rangle}
%                    math symbols
\def\a{\alpha}
\def\b{\beta}
\def\d{\delta}
\def\D{\Delta}
\def\eps{\epsilon}
\def\vareps{\varepsilon}
\def\g{\gamma}
\def\G{\Gamma}
\def\grad{\nabla}
\def\h{{1\over 2}}
\def\l{\lambda}
\def\L{\Lambda}
\def\m{\mu}
\def\n{\nu}
\def\o{\over}
\def\om{\omega}
\def\O{\Omega}
\def\p{\phi}
\def\P{\Phi}
\def\pa{\partial}
\def\pr{\prime}
\def\ra{\rightarrow}
\def\s{\sigma}
\def\S{\Sigma}
\def\t{\tau}
\def\th{\theta}
\def\Th{\Theta}
\def\ti{\tilde}
\def\wti{\widetilde}
\def\jc{J^C}
\def\jd{J^D}
\def\cw{{\cal W}}
\def\faa{Fa\'a di Bruno~}
\def\bj{{\bar J }}
\def\bp{{\bar \p}}
\def\veps{\varepsilon}
\def\tj{\ti J}
\def\tp{\ti \p}
\def\cj{{\cal J}}
\def\cbj{{\bar{\cal J}}}
\def\hb{{\hat B}}
\def\hq{{\hat Q}}
%
%%%                    macros for Lie algebras
\def\lie{{\cal G}}
\def\dlie{{\cal G}^{\ast}}
\def\elie{{\widetilde \lie}}
\def\edlie{{\elie}^{\ast}}
\def\hlie{{\cal H}}
\def\wlie{{\widetilde \lie}}
\def\sw{w_{\infty}}
\def\bw{W_{\infty}}
\def\osw{w_{1+ \infty}}
\def\obw{W_{1+\infty}}

%       fake blackboard bold macros for reals, complex, etc.
\def\rlx{\relax\leavevmode}
\def\inbar{\vrule height1.5ex width.4pt depth0pt}
\def\IZ{\rlx\hbox{\sf Z\kern-.4em Z}}
\def\IR{\rlx\hbox{\rm I\kern-.18em R}}
%
%
%               This defines remark, proposition etc.
\def\mark{\noindent{\bf Remark.}\quad}
\def\prop{\noindent{\bf Proposition.}\quad}
\def\gd{Gelfand-Dickey~}
\newcommand{\nit}{\noindent}
\newcommand{\ct}[1]{\cite{#1}}
\newcommand{\bi}[1]{\bibitem{#1}}
%
%       THIS DEFINES THE JOURNAL CITATIONS
%
\def\PRL#1#2#3{{\sl Phys. Rev. Lett.} {\bf#1} (#2) #3}
\def\NPB#1#2#3{{\sl Nucl. Phys.} {\bf B#1} (#2) #3}
\def\NPBFS#1#2#3#4{{\sl Nucl. Phys.} {\bf B#2} [FS#1] (#3) #4}
\def\CMP#1#2#3{{\sl Comm. Math. Phys.} {\bf #1} (#2) #3}
\def\PRD#1#2#3{{\sl Phys. Rev.} {\bf D#1} (#2) #3}
\def\PLA#1#2#3{{\sl Phys. Lett.} {\bf #1A} (#2) #3}
\def\PLB#1#2#3{{\sl Phys. Lett.} {\bf #1B} (#2) #3}
\def\JMP#1#2#3{{\sl J. Math. Phys.} {\bf #1} (#2) #3}
\def\JMM#1#2#3{{\sl J. Math. Mech.} {\bf #1} (#2) #3}
\def\PTP#1#2#3{{\sl Prog. Theor. Phys.} {\bf #1} (#2) #3}
\def\SPTP#1#2#3{{\sl Suppl. Prog. Theor. Phys.} {\bf #1} (#2) #3}
\def\AoP#1#2#3{{\sl Ann. of Phys.} {\bf #1} (#2) #3}
\def\PNAS#1#2#3{{\sl Proc. Natl. Acad. Sci. USA} {\bf #1} (#2) #3}
\def\RMP#1#2#3{{\sl Rev. Mod. Phys.} {\bf #1} (#2) #3}
\def\PR#1#2#3{{\sl Phys. Reports} {\bf #1} (#2) #3}
\def\AoM#1#2#3{{\sl Ann. of Math.} {\bf #1} (#2) #3}
\def\UMN#1#2#3{{\sl Usp. Mat. Nauk} {\bf #1} (#2) #3}
\def\RMS#1#2#3{{\sl Russian Math Surveys} {\bf #1} (#2) #3}
\def\FAP#1#2#3{{\sl Funkt. Anal. Prilozheniya} {\bf #1} (#2) #3}
\def\FAaIA#1#2#3{{\sl Functional Analysis and Its Application} {\bf #1} (#2)
#3}
\def\BSMF#1#2#3{{\sl Bull. Soc. Mat. France} {\bf #1} (#2) #3}
\def\BAMS#1#2#3{{\sl Bull. Am. Math. Soc.} {\bf #1} (#2) #3}
\def\TAMS#1#2#3{{\sl Trans. Am. Math. Soc.} {\bf #1} (#2) #3}
\def\AIHP#1#2#3{{\sl Ann. Inst. Henri Poincar\'e} {\bf #1} (#2) #3}
\def\AIF#1#2#3#4{{\sl Ann. Inst. Fourier} {\bf #1,#2} (#3) #4}
\def\PAMS#1#2#3{{\sl Proc. Am. Math. Soc.} {\bf #1} (#2) #3}
\def\CMJ#1#2#3{{\sl Czechosl. Math. J.} {\bf #1} (#2) #3}
\def\CompM#1#2#3{{\sl Compositio Math.} {\bf #1} (#2) #3}
\def\Invm#1#2#3{{\sl Invent. math.} {\bf #1} (#2) #3}
\def\LMP#1#2#3{{\sl Letters in Math. Phys.} {\bf #1} (#2) #3}
\def\IJMPA#1#2#3{{\sl Int. J. Mod. Phys.} {\bf A#1} (#2) #3}
\def\AdM#1#2#3{{\sl Advances in Math.} {\bf #1} (#2) #3}
\def\RMaP#1#2#3{{\sl Reports on Math. Phys.} {\bf #1} (#2) #3}
\def\IJM#1#2#3{{\sl Ill. J. Math.} {\bf #1} (#2) #3}
\def\APP#1#2#3{{\sl Acta Phys. Polon.} {\bf #1} (#2) #3}
\def\TMP#1#2#3{{\sl Theor. Mat. Phys.} {\bf #1} (#2) #3}
\def\JPA#1#2#3{{\sl J. Physics} {\bf A#1} (#2) #3}
\def\JSM#1#2#3{{\sl J. Soviet Math.} {\bf #1} (#2) #3}
\def\MPLA#1#2#3{{\sl Mod. Phys. Lett.} {\bf A#1} (#2) #3}
\def\JETP#1#2#3{{\sl Sov. Phys. JETP} {\bf #1} (#2) #3}
\def\CMH#1#2#3{{\sl Comment. Math. Helv.} {\bf #1} (#2) #3}
\def\PJAS#1#2#3{{\sl Proc. Jpn. Acad. Sci.} {\bf #1} (#2) #3}
\def\JPSJ#1#2#3{{\sl J. Phys. Soc. Jpn.} {\bf #1} (#2) #3}
\def\JETPL#1#2#3{{\sl  Sov. Phys. JETP Lett.} {\bf #1} (#2) #3}
%

%%%
\begin{titlepage}

June, 1992 \hfill{IFT-P/020/92-SAO-PAULO}

\hfill{hep-th/9206096}

\vskip .6in

\begin{center}
{\large {\bf On Two-Current Realization of KP Hierarchy}}
\end{center}

\normalsize
\vskip .4in

\begin{center}
{ H. Aratyn\footnotemark
\footnotetext{Work supported in part by U.S. Department of Energy,
contract DE-FG02-84ER40173 and by NSF, grant no. INT-9015799}}

\par \vskip .1in \noindent
Department of Physics \\
University of Illinois at Chicago\\
Box 4348, Chicago, Illinois 60680\\
\par \vskip .3in

\end{center}

\begin{center}
{L.A. Ferreira\footnotemark
\footnotetext{Work supported in part by CNPq}}, J.F. Gomes$^{\,2}$ and
A.H. Zimerman$^{\,2}$

\par \vskip .1in \noindent
Instituto de F\'{\i}sica Te\'{o}rica-UNESP\\
Rua Pamplona 145\\
01405 S\~{a}o Paulo, Brazil
\par \vskip .3in

\end{center}

\begin{center}
{\large {\bf ABSTRACT}}\\
\end{center}
\par \vskip .3in \noindent

A simple description of the KP hierarchy and its multi-hamiltonian
structure is given in terms of two Bose currents.
A deformation scheme connecting various W-infinity algebras and relation
between two fundamental nonlinear structures are discussed.
Properties of Fa\'a di Bruno polynomials are extensively explored in
this construction.

Applications of our method are given for the Conformal Affine Toda
model, WZNW models and discrete KP approach to Toda lattice chain.

\end{titlepage}

\section{Introduction}
\setcounter{equation}{0}

The integrability properties of two-dimensional conformally invariant
models can be studied in terms of extended conformal algebras \ct{zamo}.
These $W_N$ algebras are generated by the finite dimensional higher
spin objects.
They can be realized in terms of the second \gd Hamiltonian structure
of KdV hierarchies \ct{bakascmp} (see also \ct{dickey}).
The nonlinear structure of $W_N$ disappears under the large $N$ limit
giving rise to the area preserving diffeomorphisms of 2-manifold
generating the so-called $\sw$ algebra \ct{bakasplb}.
A richer linear structure appears under generalization of KdV
to KP hierarchy with the first \gd bracket \ct{watanabe}.
By this procedure one obtains an infinite dimensional Lie algebra $\bw$
\ct{wu91}.
This algebra appeared in different settings in the literature, various
isomorphic examples can be found e.g. in \ct{moyal}, \ct{pope},
\ct{radul} and \ct{khesin}.
The importance of KP hierarchy has increased recently due to its connection
to $2d$-gravity coupled to matter \ct{2d}.

In this paper we emphasize the algebraic structure of the KP hierarchy,
in terms of two Bose currents.
We show that such currents provide an unified framework for realizing
several  algebraic structures appearing in KP hierarchy, WZNW models,
Toda theories, one-matrix models and other theories possessing
W-infinity algebras.
We start with $\obw$ and its truncations (by removing lower
spin generators) all described by family of linear Watanabe-like
\ct{watanabe} structures connected with the linear parts
of \gd brackets within the KP hierarchy.
These are determined by
\be
\O^{(r)}_{n,m}\(u(x), h \) \equiv -\sum_{k=0}^{n+r} (-1)^k h^{k-1}
{n+r \choose k} u_{n+m+r-k} (x) D^k_x +
\sum_{k=0}^{m+r} h^{k-1}{m+r \choose k} D^k_x u_{n+m+r-k} (x) \phantom{aa}
\lab{hwata}
\ee
The constant $h$ is a deformation parameter responsible for extending
$\sw \to \bw$. The index $r$ denotes the level of truncation of the
underlying $\obw$.

Next we turn to the nonlinear extension of $\bw$ algebra denoted
by ${\hat W}_{\infty}$ \ct{wu92a,wu92b}.
Within the KP hierarchy it arises from the second \gd bracket and
can be governed by an extra deformation parameter $c$.
We can summarize the deformations involved in our construction in the
following simple way:
\be
\sw \stackrel{h}{\longrightarrow} \bw  \stackrel{c}{\longrightarrow}
{\hat W}_{\infty}
\lab{diagram}
\ee

In \ct{deform} the first deformation in \rf{diagram} was performed by changing
the basis from monomials in the current $J$ to $h$ dependent \faa polynomials
\ct{faa} in $J$ and its derivatives.
Within a KP hierarchy $h$ appears as one of the central elements of the
second \gd bracket.
There are two ways of generating the nonlinear structure in ${\hat
W}_{\infty}$.
One is connected with the central element in the second \gd bracket.
Alternatively one can create it by change of basis, while still working with
the first bracket.

The paper is organized as follows.
The basics of KP formalism and the \gd brackets are given in Section 2.

Section 3 provides a detailed discussion of our method to realize KP-like
algebras in terms of two currents $(\bj,J)$.
Here we make an extensive use of our basis given in terms of \faa polynomials.
We find a way of producing a family of Watanabe like forms representing the
different level of truncation of the original algebra.
We classify this structure as generated by rows and columns of the
matrix made of \faa like objects within first KP hierarchy bracket.
Next we go to the second \gd bracket, which introduces a $c$ type
deformation as in of \rf{diagram}.
We prove the equivalence of Drinfeld-Sokolov bracket with the nonlinear
part of \gd second bracket for our basis.
In this way we obtain an agreement with
Lenard relations for our formalism.
An alternative way of producing nonlinear structure in terms of
the first bracket is also dicussed.

We also study multi-Hamiltonian structures directly in terms of
$(\bj,J)$ currents.
We find a closed formula for the higher Hamiltonian brackets and for
the case $c=0$ relate it to higher Watanabe forms in \rf{hwata}.

Applications of our method are given in section 4. We make connection to WZNW
models, conformal affine Toda model \ct{bb,toda,hspin} and specific
construction of KP hierarchy related to Toda lattice chain \ct{alternative}.

The properties of \faa polynomials and some related technical identities are
given in the Appendices.

\subsection{Notation}

Throughout this paper we adopt the following notation:
\be
\pa^k f \equiv f^{(k)} \qquad ;\qquad
D \equiv {\pa \o \pa x }
\lab{not1}
\ee
$D={\pa /\pa x}$ acts on all the terms appearing to the right
as a derivative operator by the Leibniz rule.
$(A)_{-}$ denotes projection on the negative powers of $D$ contained in
the operator $A$, while $(A)_{+}$ denotes projection on the
positive and zero powers of $D$.
We define ${\rm res} (A)$ as a projection on the coefficient of
$D^{-1}$ in $A$.
Furthermore the KP flow parameters are denoted by $t$ in the
multicomponent notation $t = \( t_1, t_2, \ldots \)$.
Also, $\O^{(r)}_{n,m}\(u(x)\) \equiv \O^{(r)}_{n,m}\(u(x), h=1 \)  $.
In this paper we use $h=1$ unless stated otherwise.

\section{KP Preliminaries}
\setcounter{equation}{0}
Consider the Lax operator:
\be
L \equiv D +  \sum_{i=0}^{\infty} u_i (x, t) D^{-1-i}
\lab{laxop}
\ee
and the flows of KP hierarchy
\be
\partder {L} {t_r} = \lb \(L^r \)_{+} \, , \, L \rb \qquad\;\; r=1, 2, \ldots
\lab{kpflow}
\ee
The KP flow equation \rf{kpflow} admits a Hamiltonian structure meaning that
we can rewrite \rf{kpflow} for components of $L$ as
\be
\partder {u_i} {t_r} = \{ u_i \,, \, H_r \}_2= \{ u_i \,, \, H_{r+1} \}_1
\lab{uflow}
\ee
where the Hamiltonians for the KP hierarchy are $H_r = { 1 \o r} \int
{\rm res} \, L^r$ and $\{\cdot , \cdot \}_{1,2}$ denote first and second
Poisson bracket structure.
These higher bracket structures are compatible with Lenard relations
$\{ u_i \,, \, H_m \}_r = \{ u_i \,, \, H_{m-1} \}_{r+1}$.
Let us introduce two relevant bracket structures proposed by
Gelfand and Dickey \ct{dickey}.
Define $X = \sum_{i=0}^{\infty} \pa^i x_i $ and the pairing
\be
\me {L} {X} \equiv \Tr \( L X\) = \int {\rm res} \( L X\)
= \int \sum_{k=0}^{\infty} u_k x_k
\lab{trace}
\ee
The Gelfand-Dickey first and second bracket structures are
 defined as
\br
\{ \, \me {L} {X} \, , \, \me {L} {Y} \, \}^{GD}_1 &\equiv&
\me  { L} {\lb X\, , \, Y \rb } \lab{gd1} \\
\{ \, \me {L} {X} \, , \, \me {L} {Y} \, \}^{GD}_2 &\equiv&
\me {X} {(LY)_{-} L - L (YL)_{-}} \lab{gd2}
\er
In components the above expression for the first bracket \rf{gd1}
becomes
\br
\lefteqn{
\{ u_n (x) \, , \, u_m (y) \}^{GD}_1 = \,\O^{(r=0)}_{n,m}\(u(x)\)
\;\d (x-y)} \lab{gd1comp}\\
\O^{(r)}_{n,m}\(u(x)\) &\equiv& -\sum_{k=0}^{n+r} (-1)^k
{n+r \choose k} u_{n+m+r-k} (x) D^k_x +
\sum_{k=0}^{m+r} {m+r \choose k} D^k_x u_{n+m+r-k} (x) \phantom{aa}
\lab{wata}
\er
where we have introduced the Watanabe Hamiltonian structure
$\O^{(0)}_{n,m}$ \ct{watanabe}.
The generalizations of \rf{wata} given by
$\O^{(r)}_{n,m}$ with $r>0$ will reproduce the linear part of
several higher brackets.
Note, that one obtains $\O^{(r)}_{n,m}$ from $\O^{(0)}_{n,m}$ by a
simple shifting $u_l \to {\ti u}_l = u_{l+r}$, meaning that if $u_n$'s
satisfy \rf{gd1comp} then we will obtain $ \{ {\ti u}_n (x) \, , \,
{\ti u}_m (y) \}^{GD}_1 =  \O^{(r)}_{n,m}\({\ti u}(x)\)\, \d (x-y)$.
We will summarize this relation in the following way
\be
u_l \to {\ti u}_l = u_{l+r} \qquad;\qquad
\O^{(r)}_{n,m} \({\ti u}(x)\) = \O^{(0)}_{n+r,m+r} \(u (x)\)
\lab{shift}
\ee
The first \gd bracket structure is isomorphic to the $W_{1+\infty}$
algebra \ct{pope,wu91}.
The existence of higher structures $\O^{(r)}_{n,m}$ is therefore
related to truncation of original $W_{1+\infty}$ by removing lower
spin generators.

Higher Watanabe structures $\O^{(r)}_{n,m}$
have a nice geometric realization
in terms of functions $e^{ipx} y^{n+r}$ on a cylinder $S^1 \times \IR$ with
a Lie algebra structure given by \ct{radul}
\be
\lb f (x,y) \, ,\, g (x,y) \rb = \sum_{k \geq 1} (-h )^{k-1} {1 \o k !} \(
{\pa^k f \o \pa x^k} {\pa^k g \o \pa y^k}
-{\pa^k f \o \pa y^k} {\pa^k g \o \pa x^k} \)
\lab{symbol}
\ee
which is a commutator $\lb f , g \rb = f \circ g - g \circ f$
with respect to the product of symbols:
\be
f (x,y) \circ g (x,y) = \sum_{k \geq 0} (-h )^{k-1} {1 \o k !}
{\pa^k f \o \pa x^k} {\pa^k g \o \pa y^k}
\lab{product}
\ee
The connection becomes transparent by noticing that the mapping
\ct{deform} $W_n^p \leftrightarrow e^{ipx} y^{n+r}$ where
$W_n (x) = \sum_{p} W_n^p e^{ipx}$ is an isomorphism between
generalizations of \rf{gd1comp} with $\O^{(r)}_{n,m}$ (realizations of
such structures will be given below)
and the algebra of $e^{ipx} y^{n+r}$  under the
bracket \rf{symbol}.

For the second bracket one obtains from \rf{gd2} by direct calculation
(see for instance \ct{das}):
\be
\{ u_n (x) \, , \, u_m (y) \}^{GD}_2 =  \O^{(1)}_{n,m}\(u(x)\)\,
\d (x-y) + \{ u_n (x) \, , \, u_m (y) \}^{GD}_2 \big\v_{\rm nonlinear}
\lab{gd2comp}
\ee
with the nonlinear part given by
\br \lefteqn{
\{ u_n (x) \, , \, u_m (y) \}^{GD}_2 \big\v_{\rm nonlinear}
= \sum_{i=0}^{m-1} \Bigg\lb \sum_{k=1}^{m-i-1} {m-i-1 \choose k} u_i (x)
D^k_x u_{m+n-i-k-1} (x)        }\nonu \\
&-&\sum_{k=1}^{n} (-1)^k {n \choose k} u_{n+i-k} (x) D^k_x
u_{m-i-1} (x) \Bigg\rb \, \d (x-y)  \lab{nonligd2}\\
&-&\sum_{i=0}^{m-1} \sum_{k=0}^{n} \sum_{l =1}^{m-i-1} (-1)^k {n \choose k}
{m-i-1 \choose l} u_{n+i-k}(x) D^{k+l}_x u_{m-i-l-1} (x) \d (x-y) \nonu
\er
So far we had only defined \gd bracket structures without imposing that they
reproduce the Hamiltonian structure corresponding to KP hierarchy
flow equation \rf{kpflow}. Taking this into account one realizes
\ct{das,wu92a} that a further structure is required. Following
\ct{itzykson} we will call this structure a Drinfeld-Sokolov (DS)
bracket \ct{ds} and define it as
\br
\{ \, \me {L} {X} \, , \, \me {L} {Y} \, \}^{DS} &=&
\int dx dy \biggl( \int^x d x^{\pr} {\rm res} \lb L\, , \, X \rb \biggr) {\rm
res} \lb L \, , \, Y\rb \lab{ds} \\
\{ u_n (x) \, , \, u_m (y) \}^{DS} &= &
-\int dx_1 dx_2 \{ u_n (x) \, , \, u_0 (x_1) \}_1^{GD}
\(-\eps (x_1 - x_2) \) \{ u_0 (x_2) \, , \, u_m (y) \}_1^{GD}
\nonu\\
&=& - \sum_{i=0}^{n-1} \sum_{j =0}^{m-1} (-1)^{n-i} {n \choose i}
{m \choose j} u_i (x) D^{n+m-i-j-1}_x u_j (x) \, \d (x-y) \lab{dscomp}
\er
One notices from \rf{dscomp} that the DS bracket satisfies properties
of derivation.
Following Drinfeld-Sokolov \ct{ds,itzykson} we call two Poisson brackets
coordinated if any linear combination of them is the bracket itself
(i.e. is antisymmetric and satisfies Jacobi identity).
In fact DS bracket is coordinated with both $\O^{(0)}$ and $\O^{(1)}$ linear
bracket structures.
The second Hamiltonian structure compatible with Lenard relations is
given by \ct{das}:
\be
\{ u_n (x) \, , \, u_m (y) \}_2 = \{ u_n (x) \, , \, u_m (y) \}_2^{GD}
+ \{ u_n (x) \, , \, u_m (y) \}^{DS}
\lab{2bracket}
\ee

\section{A Bose Construction of KP Hierarchy and Fa\'a di Bruno
Polynomials}
\setcounter{equation}{0}

In this section we will construct KP hierarchy in terms of a pair of
Bose fields $J$ and $\bj$.
Based on our earlier work in \ct{deform} we propose the Lax components $W_n$
as:
\be
W_n (x) \equiv (-1)^{n}\, \bj (x) \, P_n (J(x))
\lab{wnpj}
\ee
given in terms of \faa polynomials:
\be
P_n (J) = (D +J)^n \cdot 1 = \exp(-\p) \pa^n \exp (\p) \quad;\quad
\p^{\pr} = J
\lab{defpn}
\ee
We associate to \rf{wnpj} the Lax operator given by:
\be
L = D \; +\; \bj {1 \o D + J} = D \; + \; \sum_{n=0}^{\infty}
W_n D^{-1-n}
\lab{sujinho}
\ee

{}From the definition \rf{defpn} we find a recurrence relation
\be
\pa P_n = P_{n+1} - J P_n
\lab{recc}
\ee
which could be used to calculate lowest order polynomials.
Several other useful properties of \faa polynomials are listed
in Appendix A.

\subsection{First KP Hierarchy Structure}
\lab{firstkp}

In this subsection $J$ and $\bj$ are considered as canonical
variables with commutation relation
\br
\{ \bj (x) \, , \, J (y) \}_1 &=& - \d^{\pr} (x-y) \lab{1jbarj}\\
\{ \bj (x) \, , \, \bj (y) \}_1 &= &  \{ J (x) \, , \, J (y) \}_1 = 0
\nonu
\er
which leads to
\be
\lcurl \bj (x) \, , \, \exp \( \pm \p (y)\) \rcurl_1 = \mp \d(x-y)
\exp \( \pm \p (y)\)
\lab{1jbarphi}
\ee
The advantage of using the exponential representation \rf{defpn}
is that it makes relatively easy to calculate brackets
between generators $W_n = (-1)^{n} \bj e^{ - \p}\pa^n e^{ \p}$:
\br
\lefteqn{
\lcurl W_n (x) \, ,\, W_m (y) \rcurl_1 =
(-1)^{n+m}\bj (y) e^{-\p (y)}\pa^m_y\( \d (x-y) \pa^n_y
e^{\p (y)}\) }\nonu \\
&-&(-1)^{n+m} \bj (x) e^{-\p (x)}\pa^n_x \(\d (x-y) \pa^m_x
e^{\p (x)}\) \nonu \\
&=& \sum_{k=0}^{m} (-1)^{n+m+k} {m \choose k} \bj (y) e^{ -\p (y)}
\pa^{m+n-k}_y e^{\p (y)}\,\pa_x^{(k)} \d (x -y) \nonu\\
&-& \sum_{k=0}^{n} (-1)^{n+m+k} {n\choose k} \bj (x) e^{ -\p (x)}
\pa^{m+n-k}_x e^{ \p (x)}\,\pa_y^{(k)} \d (x -y)
\lab{wnwm}
\er
One easily recognizes in \rf{wnwm} the first \gd structure written in
the Watanabe form:
\be
\{W_n (x) \, , \, W_m (y) \}_1 = \O_{nm}^{(0)} (W(x))\, \d (x-y)
\lab{wnwata}
\ee

We now define a family of related generators
\be
W_n^{(k)} = (-1)^{n+k} \( (D- J)^k \bj \)  P_{n} (J)  \qquad;\qquad
n,k=0,1,2,\ldots
\lab{genbasis}
\ee
Note, that $W_n^{(0)} = W_n$.
We now establish, with help of \rf{recc}, the following recursive
relation:
\be
W_n^{(k+1)} = - W_{n+1}^{(k)} - \pa W_n^{(k)} \qquad;\qquad n,k=0,1,2,
\ldots
\lab{k2kplusone}
\ee
The above relation allows the transparent interpretation
in terms of the infinite matrix
\be
{\cal W} = \left (\begin{array}{cccc}
W^{(0)}_0 & W^{(1)}_0 & W^{(2)}_0 &\cdots  \\
W^{(0)}_1 & W^{(1)}_1 & W^{(2)}_1 &\cdots \\
W^{(0)}_2 & W^{(1)}_2 & W^{(2)}_2 &\cdots  \\
\vdots & \vdots & \vdots &\, \end{array}
\right )
\lab{wmatrix}
\ee
In such matrix notation the recurrence relation \rf{k2kplusone} reads
as
\be
\pa {\cal W} = - I_{+} {\cal W} - {\cal W} I_{-}
\lab{matrec}
\ee
where we used the infinite raising and lowering matrices
\be
I_{-}=\left (\begin{array}{ccccc}
0 & 0 & 0 & 0 & \cdots \\
1 & 0 & 0 & 0 & \cdots \\
0 & 1 & 0 & 0 & \cdots \\
0 & 0 & 1 & 0 & \cdots\\
\vdots & \vdots & \,&\ddots &\ddots \end{array}
\right ) \;\;\; , \;\;\;
I_{+} =\left (\begin{array}{ccccc}
0 & 1 & 0  & 0 & \cdots \\
0 & 0 & 1  & 0 & \cdots \\
0 & 0 & 0  & 1 & \, \\
0 & 0 & 0  & 0 & \ddots \\
\vdots & \vdots&\vdots & \vdots  & \ddots \end{array}
\right )
\lab{iplus}
\ee
Note that the top row of ${\cal W}$ matrix \rf{wmatrix} consisting
of elements
\be
W_0^{(k)} = (-1)^{k} (D - J)^k \bj \qquad;\qquad
k=0,1,2,\ldots
\lab{rowbasis}
\ee
is generated by its own Lax operator
\be
L = D \; +\; {1 \o D - J} \bj = D \; + \; \sum_{n=0}^{\infty}
W_0^{(k)} D^{-1-k}
\lab{conjlax}
\ee
It is natural to introduce in this setting the concept of adjoint
operation, which maps the matrix \rf{wmatrix} into its transposed
(with rows and columns interchanged).
In operator language we notice from \rf{genbasis} and definition
\rf{defpn} that adjoint operation is defined by swapping each $D+J$
to the right of $\bj$ with $D-J$ to the left of $\bj$ and vice versa.
This concept will be useful when we discuss below the algebra involving
all generators defined in \rf{genbasis}.

We now show that the generators defined in \rf{genbasis}
satisfy the closed algebra and we find its form.
In order to do that we introduce the quantities
\br
{\cal U}^{(r)}_{n,m}(x) &\equiv &(-1)^r \sum_{k=0}^{n} (-1)^k {n \choose k}
W^{(r)}_{n+m-k} D^k_x
\lab{u}\\
{\cal V}^{(r)}_{n,m}(x) &\equiv &(-1)^r \sum_{k=0}^{m} {m \choose k} D^k_x
W^{(r)}_{n+m-k}
\lab{v}
\er
In the appendix \rf{appendixb} we show that, these quantities satisfy
the following identities
\br
{\cal
U}^{(r+1)}_{n,m}(x) &=&
{\cal U}^{(r)}_{n+1,m}(x) + D_x {\cal U}^{(r)}_{n,m}(x)
\lab{id1}\\
{\cal
V}^{(r)}_{n,m+1}(x) &=&
{\cal V}^{(r)}_{n+1,m}(x) + D_x {\cal V}^{(r)}_{n,m}(x)
\lab{id2}\\
{\cal U}^{(r)}_{n+1,m}(x) \d (x-y) &=&
\lb {\cal U}^{(r)}_{n,m+1}(x) + D_y {\cal U}^{(r)}_{n,m}(x)\rb \d (x-y)
\lab{id3}\\
{\cal V}^{(r+1)}_{n,m}(x) \d (x-y) &=&
\lb {\cal V}^{(r)}_{n,m+1}(x) + D_y {\cal V}^{(r)}_{n,m}(x)\rb \d (x-y)
\lab{id4}
\er
The bracket \rf{wnwata} between the generators $W_n(x) \equiv W_n^{(0)}(x)$
can then be written as
\be
\lcurl W_n^{(0)} (x) \, ,\, W_m^{(0)} (y) \rcurl_1 = - \lb {\cal
U}^{(0)}_{n,m}(x) -  {\cal V}^{(0)}_{n,m}(x) \rb \d (x-y)
\lab{zerocomm}
\ee
The bracket between any two generators $W_n^{(r)}(x)$ can be obtained
recursively from \rf{zerocomm} by using the identities
\rf{id1}-\rf{id4} and the relation \rf{k2kplusone}. For instance,
using \rf{k2kplusone}, \rf{id1} and \rf{id2} one gets
\br
\lcurl W_n^{(1)} (x) \, ,\, W_m^{(0)} (y) \rcurl_1 &=&
-\lcurl W_{n+1}^{(0)} (x) \, ,\, W_m^{(0)} (y) \rcurl_1
-D_x \lcurl W_n^{(0)} (x) \, ,\, W_m^{(0)} (y) \rcurl_1 \nonu \\
&=& \biggl( {\cal U}^{(1)}_{n,m}(x) -  {\cal V}^{(0)}_{n,m+1}(x) \biggr)
\d (x-y) \nonu
\er
Analogously, using \rf{k2kplusone}, \rf{id3} and \rf{id4} one finds
\be
\lcurl W_n^{(0)} (x) \, ,\, W_m^{(1)} (y) \rcurl_1 = \biggl(
{\cal U}^{(0)}_{n+1,m}(x) -  {\cal V}^{(1)}_{n,m}(x) \biggr)
\d (x-y)   \nonu
\ee
By repeating the above steps we arrive at the general expression
\be
\lcurl W_n^{(r)} (x) \, ,\, W_m^{(s)} (y) \rcurl_1 =
- (-1)^{r+s} \biggl( {\cal U}^{(r)}_{n+s,m}(x) -  {\cal V}^{(s)}_{n,m+r}(x)
\biggr) \d (x-y)
\lab{general}
\ee
completing the proof of the closure of the algebra.

Notice that the generators $W$'s belonging to the same column of
the matrix \rf{wmatrix}, i.e. $r=s$, constitute a closed subalgebra.
In fact it is isomorphic to a truncated Watanabe type algebra
via \rf{shift}.
\be
\lcurl W_n^{(r)} (x) \, ,\, W_m^{(r)} (y) \rcurl_1 =
(-1)^r \Omega_{n,m}^{(r)} (W^{(r)}) \d (x-y)
\lab{truncatedcolumn}
\ee
for $n,m= 0 , 1, \ldots$.

The generators appearing on a given row of the matrix \rf{wmatrix} also
constitute a closed subalgebra. In appendix \rf{appendixc} we show that
\be
\lcurl W_n^{(r)} (x) \, ,\, W_n^{(s)} (y) \rcurl_1 = \biggl( {\tilde {\cal
U}}^{r+n,s}_n - {\tilde {\cal V}}^{r,n+s}_n \biggr) \d (x-y)
\lab{linealgebra}
\ee
where we have defined
\br
{\tilde {\cal U}}^{r,s}_n (x) &\equiv & (-1)^n \sum_{k=0}^r (-1)^k {r \choose
k}
W^{(r+s-k)}_n (x) D_x^k
\lab{utilde}\\
{\tilde {\cal V}}^{r,s}_n (x) &\equiv & (-1)^n \sum_{k=0}^s {s \choose k}
D_x^k W^{(r+s-k)}_n (x)
\lab{vtilde}
\er
The subalgebra \rf{linealgebra} is also a truncated Watanabe type
algebra:
\be
\lcurl W_n^{(r)} (x) \, ,\, W_n^{(s)} (y) \rcurl_1 =
- (-1)^n \Omega_{r,s}^{(n)} (W_n) \d (x-y)
\lab{truncatedline}
\ee
for $r,s=0,1,\dots$.

\subsection{Second KP Hierarchy Structure}

In this subsection we will show how to generate the second bracket
structure from the representation given by \rf{wnpj}.
This time the algebra of $J$ and $\bj$ will be defined to be
\br
\{ \bj (x) \, , \, J (y) \}_2 &=& J(x) \d^{\pr} (x-y) - h \d^{\pr\pr} (x-y)
\nonu\\
\{ \bj (x) \, , \, \bj (y) \}_2 &= &  2 \bj (x) \d^{\pr} (x-y) +\bj^{\pr}
(x) \d (x-y) \lab{2jbarj}\\
\{ J (x) \, , \, J (y) \}_2 &=& c \, \d^{\pr} (x-y) \nonu
\er
where constants $h$ and $c$ can be interpreted as
independent deformation parameters, see next subsection for details.
Here we take $h=1$.
Recalling from \rf{wnpj} that $W_0=\bj$, $W_1 = - \bj J$ and
$W_2= \bj (J^{\pr} + J^2)$ one can easily check that \rf{2jbarj} is
a unique structure for $(\bj,J)$ leading to \rf{2bracket} for three
lowest brackets with $c=2$.
{}From \rf{2jbarj} we derive
\be
\lcurl \bj (x) \, , \, \exp \( \pm \p (y)\) \rcurl_2 = \mp \d(x-y)
\pa \exp \( \pm \p (y)\) \pm \d^{\pr} (x-y) \exp \( \pm \p (y)\)
\lab{2jbarphi}
\ee
repeating similar calculation as in \rf{wnwm} we get
for the linear part of $\{\cdot,\cdot\}_2$ the expected
result $\{W_n (x) \, , \, W_m (y) \}_2 \v_{\rm linear} =
\O_{nm}^{(1)} (W(x))\, \d (x-y)$.
To calculate the nonlinear part of the bracket we will use the
exponentional representation of \faa polynomials given in \rf{defpn}.
We first observe that
\be
\{ \p (x) \, , \, \p (y) \}_2 = - c \, \veps (x-y) \qquad;\qquad
\p^{\pr}(x) = J (x)
\lab{phiphi}
\ee
from which the direct calculation yields
\br
\lefteqn{
\lcurl P_n (x) \, ,\, P_m (y) \rcurl_2 \big\v_{\rm nonlinear} = - c
\Bigg\lb \sum_{l=0}^n \sum_{p=0}^m {n \choose l} {m \choose p}
P_{n-l} (x) P_{m-p} (y) \pa^l_x \pa^p_y }\nonu\\
&-& P_{n} (x) \sum_{l=0}^m {m\choose l}
P_{m-l}(y)\pa^l_y -P_{m} (y) \sum_{l=0}^n {n\choose l} P_{n-l}(x)
\pa^l_x \nonu \\
&+&P_{n} (x) P_{m}(y) \Bigg\rb \veps (x-y)
\lab{nonlinpart}
\er
where we wrote for brevity $P_n \( J(x) \) = P_n (x)$.
An important point is that the pure $\veps (x-y)$ terms
cancel out leaving only delta functions and their derivatives
in the following expression:
\be
\lcurl P_{n} (x) \, ,\, P_{n} (y) \rcurl_2^{\rm nonlinear}
= -c \Bigg\lb \sum_{l=1}^n \sum_{p=1}^m (-1)^{p} {n \choose l}
{m \choose p} P_{n-l}(x) P_{m-p} (y) \pa^{l+p-1}_x \Bigg\rb \d (x-y)
\lab{pnpmnonlin}
\ee
We obtain therefore the total second bracket for the generators in
\rf{wnpj} as the sum of linear and nonlinear terms (after
a change of variables $n-l=i$, $m-p=j$ in \rf{pnpmnonlin}):
\br
\lefteqn{
\lcurl W_{n}(x) \, ,\, W_{m} (y) \rcurl_2 = \O_{nm}^{(1)}\(W(x)\)\,
\d (x-y) }\lab{third}\\
&-& c \Bigg\lb \sum_{i=0}^{n-1} \sum_{j=0}^{m-1} (-1)^{n-i}
{n \choose i} {m \choose j} W_{i}(x) D^{n+m-i-j-1}_x W_{j} (x)
\Bigg\rb \d (x-y)   \nonu
\er
for $n,m \geq 0$.

This result appears at first to be surprising since we recognize in
\rf{third} only the DS structure from \rf{dscomp} multiplied by $c$
while the nonlinear part of the second \gd bracket from \rf{gd2comp}
appears to be missing.
This rises the question whether agreement with second Hamiltonian
structure of \rf{2bracket} encountered above at the lowest levels
holds at the higher level for our basis \rf{wnpj}.
The following proposition shows that this is indeed the case.

\prop The following identity:
\be
\{ W_n (x) \, , \, W_m (y) \}^{GD}_2 \bigg\v_{\rm nonlinear}
= \{ W_n (x) \, , \, W_m (y) \}^{DS}
\lab{2=3}
\ee
involving the nonlinear part of the second \gd bracket \rf{nonligd2}
and the DS structure \rf{dscomp} is valid for the generators
$W_n (x) = (-1)^{n}\, \bj (x) \, P_n (J(x))$.

Let us first recall from \rf{gd2comp} and \rf{dscomp}
that identity \rf{2=3} reads in components:
\br \lefteqn{
\sum_{i=0}^{m-1} \Bigg\lb \sum_{k=1}^{m-i-1} {m-i-1 \choose k} W_i
D_x^k W_{m+n-i-k-1}
-\sum_{k=1}^{n} (-1)^k {n \choose k} W_{n+i-k} D_x^k W_{m-i-1} \Bigg\rb
\, \d (x-y)}  \nonu   \\
&-&\sum_{i=0}^{m-1} \sum_{k=0}^{n} \sum_{l =1}^{m-i-1} (-1)^k {n \choose k}
{m-i-1 \choose l} W_{n+i-k} D_x^{k+l} W_{m-i-l-1} \d (x-y)
\phantom{aaaaaaa} \lab{2=3comp}\\
&=& - \sum_{i=0}^{n-1} \sum_{j =0}^{m-1} (-1)^{n-i} {n \choose i}
{m \choose j} W_i D_x^{n+m-i-j-1} W_j\, \d (x-y) \nonu
\er
with all $W_n$'s taken at $x$.
We will prove this identity by induction. First, note that both sides of
\rf{2=3comp} are zero for $n=0$ and arbitrary $m$. A short calculation
shows that they are also equal for $n=1$ and arbitrary $m$:
\be
\{ W_1 (x) \, , \, W_m (y) \}^{GD}_2 \bigg\v_{\rm nonlinear}
= \{ W_1 (x) \, , \, W_m (y) \}^{DS} =
\sum_{k=1}^m { m \choose k} W_0 (x) D^k_x W_{m-k} (x) \d (x-y)
\lab{n=1m}
\ee
Note now that since one can factorize out $\bj (x)$ and $\bj (y)$ on both
sides of \rf{2=3comp} and \rf{n=1m}
we can substitute there $W_l $ by $(-1)^l P_l$.
Hence the proof requires showing that
\be
\{ P_n (x) \, , \, P_m (y) \}^{GD}_2 \bigg\v_{\rm nonlinear}
= \{ P_n (x) \, , \, P_m (y) \}^{DS}
\lab{induction}
\ee
We make an induction assumption that \rf{induction} holds for some
fixed $n$ and arbitrary $m$.
Let us now show that this identity is also true for $n+1$ with
arbitrary $m$.
Recall from \rf{recc} that $ P_{n+1}=  - \pa P_n + P_1 P_n$.
The rest of the proof follows now
easily from \rf{n=1m}, induction assumption \rf{induction} and
the derivation property of both brackets.

As a consequence of the above proposition we are able to rewrite
relation \rf{third} for $c=2$ as
\be
\lcurl W_{n}(x) \, ,\, W_{m} (y) \rcurl_2
= \{ W_n (x) \, , \, W_m (y) \}_2^{GD} + \{ W_n (x) \, , \, W_m (y) \}^{DS}
\lab{w2bracket}
\ee
where on the right hand side of \rf{w2bracket} we had split nonlinear
part of \rf{third} equally between $\{\cdot,\cdot\}_2^{GD} \v_{\rm
nonlinear}$ and $\{\cdot,\cdot\}^{DS}$.

\subsection{Nonlinear Structure from the First Bracket}

In this subsection we present a special realization of \rf{diagram} in terms
of the first bracket structure \rf{1jbarj} by changing the basis.

Our initial basis is
\be
w_n = (-1)^n {\bar J} J^n
\ee
for $n=0,1,2...$ , which generates the area preserving diffeomorphism algebra
$w_{1+\infty}$:
\be
\lcurl w_n(x)  \, , \, w_m(y) \rcurl_1 = \( n w_{m+n-1}(x) D_x + m D_x
w_{m+n-1}(x) \) \d (x-y)
\ee
In order to introduce a deformation parameter $h$ we replace ${\bar J}$ by
${\bar{\cal J}} \equiv (h \pa - J ){\bar J}$. It satisfies
\be
\lcurl \cbj (x) \, , \, \cbj (y) \rcurl_1 = 2 \cbj(x) \d^{\pr}(x-y) +
\cbj^{\pr}(x) \d (x-y)
\ee
We also have
\be
\lcurl \cbj (x) \, , \, J(y) \rcurl_1 = J(x) \d^{\pr}(x-y) - h \d^{\pr\pr}(x-y)
\ee
We then see we have generated \rf{2jbarj} with $c=0$ out of \rf{1jbarj}
by changing the basis. In the basis $(\cbj,J)$ we define
\be
V_n \equiv (-1)^n \cbj P_n(J)
\ee
satisfying $W_{\infty}$ algebra with the Watanabe structure
$\O^{(1)}_{n,m}(V)$.

In order to introduce the $c \ne 0$ into the algebra let us now define
$\cj \equiv J - {c \over 2} \bj $ satisfying
\be
\lcurl \cj (x) \, , \, \cj (y) \rcurl_1 = c \d^{\pr}(x-y)
\ee
Further we obtain
\be
\lcurl \cbj (x) \, , \, \cj (y) \rcurl_1 = \cj (x) \d^{\pr}(x-y) - h
\d^{\pr\pr}(x-y)
\ee
In the basis $(\cbj,\cj)$ we now can define generators of the
${\hat W}_{\infty}$ algebra
\be
{\hat V}_n (x) = (-1)^n \cbj P_n(\cj )
\ee
Comparing with analysis of section 3.2 we immediately conclude that
${\hat V}_n $ satisfies the second Hamiltonian structure
\rf{w2bracket}.

In reference \ct{wu92b} there exists proposal similar in spirit to the
above discussion.
Despite the technical differences we show in detail how to incorporate
it in our framework
Start by defining
\br
\cw_n &=& \sum_{k=0}^n (-1)^{k-1} {n \choose k} W_{n-k+1} P_{k} (\tj)
= \tj \sum_{k=0}^n (-1)^{k-1} {n \choose k} e^{-\p} \pa^{n-k}
\(\pa e^{\p}\) e^{-\tp} \pa^{k} e^{\tp} \nonu\\
&=& (-1)^n \tj e^{-\p-\tp } \pa^{n} \( (\pa e^{\p}) e^{\tp} \)
= (-1)^n \tj e^{-\Phi } \pa^{n} \(J e^{\Phi}\) \lab{convol}
\er
where we have introduced the current $\tj$ satisfying
$\{ \tj (x) , J (y) \} = \d^{\pr} (x-y)$ with remaining brackets being
zero. Also $\tp$ satisfies $\tp^{\pr} = \tj$ and $\Phi= \tp + \p$.
The expression \rf{convol} can also be rewritten as
\be
\cw_n = (-1)^n \tj e^{-\Phi } D^{n} e^{\Phi} J =
(-1)^n \tj \( D + \tj + J \)^n J
\lab{cwn}
\ee
Accordingly the Lax representation becomes:
\be
L = D + \tj { 1 \o D + \tj + J} J = D + \sum_{n=0}^{\infty} \cw_n
D^{-n-1}
\lab{nlaxa}
\ee
To calculate the algebra satisfied by $\cw_n$
it is useful to introduce a vertex function representation.
We recall first that from equations \rf{ephi} and \rf{schur} from
Appendix A we have
\be
e^{\D_{\eps} \Phi (x)}
= \sum_{n=0}^{\infty} P_n\( \tj +J \) {\eps^n \o n!}
\lab{bephi}
\ee
where we have introduced for brevity a notation
$\D_{\eps} \Phi (x)= \Phi (x+\eps ) - \Phi (x)$.
In this notation we can write:
\be
\tj (x) J (x + \eps) e^{\D_{\eps} \Phi (x)}=
\sum_{n=0}^{\infty} (-1)^n \cw_n (x) {\eps^n \o n!}
\lab{cweps}
\ee

Hence we will be interested in calculating brackets using the generating
functions from \rf{cweps}.
The relevant bracket is calculated in Appendix \rf{appendixd} and
is given by:
\br
\lefteqn{
\lcurl \cw_{n}(x) \, ,\, \cw_{m} (y) \rcurl =
\O^{(1)}_{nm} \(\cw (x)\) \, \d (x-y) }\lab{wuyu}\\
&-& 2 \Biggl( \sum_{i=0}^{n-1} \sum_{j=0}^{m-1} (-1)^{n-i} {n \choose i}
{m \choose j} \cw_{i}(x) D^{n+m-i-j-1}_x \cw_{j} (x)   \Biggr) \d (x-y)
\nonu
\er
As before the result of direct calculation of brackets resulted only in
DS structure as a nonlinear part.
We will now show that also in this case there is an equivalence with
the \gd second bracket.
The proof is simplified by writing the generators $\cw_n$ as
\be
\cw_n = (-1)^n \tj R_n ( \tj +J) \qquad;\qquad R_n = (-1)^n \( D + J +
\tj \)^n J
\lab{cwnrn}
\ee
with $R_0 = J$ and $R_1 = - J^{\pr} - (\tj + J ) J$.
Since we can always factorize $\tj(x) \tj(y)$ out in the bracket \rf{wuyu}
we will really be interested in proving
\be
\{ R_n (x) \, , \, R_m (y) \}^{GD}_2 \bigg\v_{\rm nonlinear}
= \{ R_n (x) \, , \, R_m (y) \}^{DS}
\lab{rinduction}
\ee
where the detailed expressions for the left and right hand side are
given in \rf{nonligd2} and \rf{dscomp}.
One verifies easily that
\be
\{ R_0 (x) \, , \, R_m (y) \}^{GD}_2 \bigg\v_{\rm nonlinear}
= \{ R_0 (x) \, , \, R_m (y) \}^{DS} = 0
\lab{r0rm}
\ee
and
\be
\{ R_1 (x) \, , \, R_m (y) \}^{GD}_2 \bigg\v_{\rm nonlinear}
= \{ R_1 (x) \, , \, R_m (y) \}^{DS}
\lab{r1rm}
\ee
Since $R_0 = J$ and $R_1 = - R_0^{\pr} - R_0^2 - \tj R_0$ we obtain
\be
\{ (\tj+J) (x) \, , \, R_m (y) \}^{GD}_2 \bigg\v_{\rm nonlinear}
= \{ (\tj+J) (x) \, , \, R_m (y) \}^{DS}
\lab{bjjrm}
\ee
This last relation makes now possible for us to use the recurrence relation
$R_{n+1} = -R_n^{\pr} -(\tj+J) R_n$ to complete the induction proof.

\subsection{KP Multi-Hamiltonian Structure in Terms of Currents $J,\bj$}

The results of subsections (3.1) and (3.2) suggest the possibility of
transfering the KP flow equation \rf{kpflow} to its Hamiltonian form
solely in terms of Bose currents $(J,\bj)$ instead of the Lax components
$u_n$'s.
Similar idea appeared in \ct{depi} in the setting of formalism
given in \ct{wu92b}.

In terms of the currents the equation \rf{uflow}
${\pa u_i}/ {\pa t_r} = \{ u_i \,, \, H_{r+2-k} \}_k$
for $k=1,2$ reads as:
\br
\partder {J} {t_r} &=& \{ J\,, \, H_r \}_2=
\{ J \, , \, H_{r+1} \}_1 \lab{jflow}\\
\partder {\bj}{t_r} &= &\{ \bj \,, \, H_r \}_2
= \{ \bj \,, \, H_{r+1} \}_1         \nonu
\er
which rewritten in a ``spinor" form becomes
\be
{ {\d J}/ {\d t_r} \choose {\d \bj}/ {\d t_r} }
= P_1 { {\d {\cal H}_{r+1} }/ {\d J} \choose
{\d {\cal H}_{r+1} }/ {\d \bj}} =
P_2 { {\d {\cal H}_{r} }/ {\d J} \choose
{\d {\cal H}_{r} }/ {\d \bj}}
\lab{sjflow}
\ee
where $P_{1,2}$ are Hamiltonian structures for the $(\bj, J)$ KP
hierarchy written in the matrix form.
In terms of $P_i$ we can express the $i$-th bracket as
\be
\{ A\, , \, B \}_i = \( {\d A \o \d J} \; {\d A \o \d \bj} \)\; P_i \;
{\d B / \d J \choose  \d B/ \d \bj }
\lab{abbra}
\ee
Equation \rf{sjflow} expresses compatibility relation between the first
and second Hamiltonian structures. We can put it in the form of
reccurence relation:
\be
{ {\d {\cal H}_{r+1} }/ {\d J} \choose
{\d {\cal H}_{r+1} }/ {\d \bj}} =
(P_1 )^{-1} P_2 { {\d {\cal H}_{r} }/ {\d J} \choose
{\d {\cal H}_{r} }/ {\d \bj}}
= \((P_1 )^{-1} P_2\)^{i-1} { {\d {\cal H}_{r+2-i} }/ {\d J} \choose
{\d {\cal H}_{r+2-i} }/ {\d \bj}}
\lab{compati}
\ee
for any $i$ such that $1 \leq i \leq r+1$.
Recalling that for the general Hamiltonian matrix structure $P_i$,
we have
\be
{ {\d J}/ {\d t_r} \choose {\d \bj}/ {\d t_r} }
= P_i \, { {\d {\cal H}_{r+2-i}}/ {\d J} \choose
{\d {\cal H}_{r+2-i}}/ {\d \bj}}
\lab{iflow}
\ee
we find the following relation between Hamiltonian matrix structures
\br
P_i &=& P_1 \((P_1 )^{-1} P_2\)^{i-1} = P_{i-1} (P_1 )^{-1} P_2
\qquad\; i\geq 1 \lab{pip1p2}\\
P_i &=& P_2 \((P_1 )^{-1} P_2\)^{i-2} \qquad\; i\geq 2\nonu
\er
Hence among the multi-Hamiltonian structures only $P_1$ and $P_2$ are
independent.
All other matrices $P_i\;,\; i=3,4,\ldots$ are related
through \rf{pip1p2} to $P_1$ and $P_2$.
{}From \rf{1jbarj} and \rf{2jbarj} we find directly
\be
P_1 = \left(\begin{array}{cc}
0 & - \pa \\
-\pa & \; 0 \end{array}
\right) \;\; , \;\;
P_2 =\left(\begin{array}{cc}
c \pa & \; \pa^2 + \pa J \\
- \pa^2 + J \pa &\; \pa \bj+ \bj \pa \end{array}
\right)
\lab{p1p2}
\ee
In particular, for $i=3$ we find $P_3 = P_2 P_1^{-1} P_2$,
which gives explicitly
\br
P_3 &=& c {\bar P}_3 + P_3 \big\v_{c=0} \lab{p3}\\
&= &
- c \left(\begin{array}{cc}
\pa J+ J \pa & \pa \bj+ \bj \pa \\
\pa \bj+ \bj \pa &  0 \end{array}
\right)
- \(\begin{array}{cc}
0   & \pa \( \pa + J \)^2 \\
\(- \pa + J \)^2 \pa & \;\, \bj \( \pa + J \)^2 - \(- \pa + J \)^2 \bj
\end{array} \)
\nonu
\er
The higher brackets will however contain nonlocal expressions
in their $c$-dependent parts.
The simplicity of $P_3 \v_{c=0} $ is maintained at the level of higher
Hamiltonian structures. In fact we have:
\be
P_n \big\v_{c=0} = (-1)^n \(\begin{array}{cc}
0   & \pa \( \pa + J \)^{n-1} \\
\(- \pa + J \)^{n-1} \pa & \;\, \bj \( \pa + J \)^{n-1}
- \(- \pa + J \)^{n-1} \bj \end{array} \)
\lab{pn}
\ee
This statement is easily proven by induction. We find from
\rf{pip1p2} that $P_{n+1} \v_{c=0}
= (P_n \v_{c=0}) P_1^{-1} (P_2 \v_{c=0})$.
Inserting the induction assumption \rf{pn} into this last equation
we complete the proof after a short calculation.
Hence for $c=0$ the corresponding $(\bj, J)$ algebra would take the
following simple form for the n-th Hamiltonian structure ($n \geq 1)$:
\br
\{ \bj (x) \, , \, J (y) \}_n \big\v_{c=0} &=& (-1)^n
\bigl( -\pa + J(x) \bigr)^{n-1} \d^{\pr} (x-y)\lab{njbarj} \\
\{ \bj (x) \, , \, \bj (y) \}_n \big\v_{c=0} &= & (-1)^n \biggl(
\bj (x) \( \pa + J (x) \)^{n-1} - \(- \pa + J(x)\)^{n-1} \bj (x)
\biggr) \d (x-y) \nonu \\
\{ J (x) \, , \, J (y) \}_n \big\v_{c=0} &=& 0 \nonu
\er
It is interesting to note that the form of the bracket
$\{ \cdot , \cdot \}_n \v_{c=0} $ in \rf{njbarj} can be reproduced by
the lower bracket with modified basis.
This construction goes as follows. Let $ \bj \to (-\pa + J) \bj =
W_0^{(1)}$. Then we find
\br
\{ W_0^{(1)} (x) \, , \, J (y) \}_{n-1} \big\v_{c=0}& =&- (-1)^n
\bigl( -\pa + J(x) \bigr)^{n-1} \d^{\pr} (x-y)\lab{n1jbarj} \\
\{ W_0^{(1)} (x) \, , \, W_0^{(1)} (y) \}_{n-1} \big\v_{c=0}& = &
- (-1)^n \biggl( W_0^{(1)} (x) \( \pa + J (x) \)^{n-1} \nonu \\
&-& \(- \pa + J(x)\)^{n-1} W_0^{(1)} (x) \biggr) \d (x-y) \nonu
\er
Hence the form of the bracket was preserved under simultaneous change of
basis and lowering the index of the bracket. Continuing this process
till we reach the first bracket structure we arrive at
$ \{ W_0^{(n-1)} (x) \, , \, W_0^{(n-1)} (y) \}_{1}$ known from
\rf{truncatedcolumn}.
Hence $\{ W_n (x) , W_m (y) \}_{r} \v_{c=0}$
will have the same functional form as $(-1)^r \{ W_n^{(r-1)} (x) ,
W_m^{(r-1)} (y) \}_{1}$.
{}From here we find, recalling equation \rf{truncatedcolumn}, that the
algebra of $W_n = (-1)^n \bj P_n (J)$ according to the r-th order
Poisson structure is given by
\be
\{ W_n (x) \, , \, W_m (y) \}_r \big\v_{c=0}
= \O^{(r-1)}_{nm} \(W(x)\) \, \d (x-y)  \lab{wnwmrth}
\ee
In the light of above result we observe that the algebra of columns (or
rows) of the matrix \rf{wmatrix} is isomorphic to the higher undeformed
algebras in \rf{wnwmrth}.
One also concludes that $c$ plays a role of a deformation parameter
responsible for deformation of the linear $\bw$ algebra to
${\hat W}_{\infty}$ algebra.

We can use the two bracket structures to construct a Lax equation as
discussed in ref. \ct{dasb}. Denoting $J^1 \equiv J$ and
$J^2 \equiv {\bar J}$, we define the $2 \times 2$ matrices
\br
S^j_i &\equiv &\( P^{-1}_1 \)_{ik} \( P_2 \)^{kj} \nonumber \\
\( U_r\)_i^j &\equiv  & {\pa \over \pa J^i}{\pa J^j \over \pa t_r}
\lab{das1}
\er
{}From the compatibility condition \rf{sjflow} and the fact that $P_r$ are
symplectic forms, and therefore closed, one gets that $S$ and $U$
constitute a Lax pair \ct{dasb}
\be
{d S  \over d t_r}=  \lb S \, , \, U_r \rb
\lab{daslax}
\ee
{}From \rf{das1} and \rf{p1p2} one easily gets
\br
S= \(
\begin{array}{cc}
\pa - \pa^{-1} J \pa & -(\bj + \pa^{-1} \bj \pa )\\
-c & -(\pa  + J)
\end{array}\)
\er
and for $r=1$, ${\cal H}_1 = \bj $ and ${\cal H}_2 = - \bj J$
\br
U_1 = \(
\begin{array}{cc}
\pa & 0\\
0 &  \pa
\end{array}\)
\er
Using these matrices one concludes that \rf{daslax} is compatible with the
equations of motion
\be
{\pa J \over \pa t_1} = J^{\pr} \, \, \, , \, \, \,
{\pa \bj \over \pa t_1} = \bj^{\pr}
\ee
One can use \rf{daslax}  to construct conserved quantities as $ \Tr S^n$
where $\Tr$ is some invariant trace form for these operators.

\section{Applications}
\setcounter{equation}{0}

\subsection{WZNW Type Models}
The ordinary WZNW model associated to a Lie group $G$ possesses two
commuting chiral copies of the current algebra:
\be
\lcurl J_a(x) \, ,\, J_b(y) \rcurl = f_{ab}^c J_c(x) \d (x-y) + k
g_{ab}
 \d^{\pr}(x-y)
\lab{ordca}
\ee
where $f_{ab}^c$ are the structure constants of the Lie algebra $\cal G$ of
$G$, and $g_{ab}$ is the Killing form of $\cal G$.
The two chiral components of the energy momentum tensor
are of the Sugawara form
\be
T(x) = \sum_{a,b=1}^{dim G} g^{ab} J_a(x) J_b(x)
\ee
where $g^{ab}$ is the inverse of the Killing form. Such tensor
satisfies the Virasoro algebra with vanishing central term
\be
\lcurl T(x) \, ,\, T(y) \rcurl = 2 T(x)  \d^{\pr}(x-y) +  T^{\pr}(x)
\d (x-y)  \lab{virsuga}
\ee
The currents are spin one primary fields
\be
\lcurl T(x) \, ,\, J_a(y) \rcurl = J_a (x)  \d^{\pr}(x-y)
\lab{primary}
\ee
Suppose now one has a self-commuting current ${\cal J}$,
$\lcurl {\cal J}(x) \, ,\, {\cal J}(y) \rcurl = 0$.
For the non compact WZNW model this current can be, for instance, the one
associated to a step operator $J(E_{\a})$ for any root $\a$ of $G$.
One then sees that the system
$(T,{\cal J})$ generates an algebra isomorphic to \rf{2jbarj} with $h=c=0$,
where
$T$ corresponds to ${\bar J}$ and $\cal J$ to $J$.
One can construct out of them the quantities  $w_n(x)
\equiv  T(x) {\cal J}^{n-2}$ satisfying the area preserving diffeomorphism
algebra, which correspond to \rf{third} for $h=c=0$, i.e.
\be
\{ w_n (x) \, , \, w_m (y) \} = \( n + m -2 \) w_{n+m-2} (x) \d^{\pr} (x - y)
+( m-1 ) \( w_{n+m-2 } (x)\)^{\pr} \d (x -y)  \phantom{......}
\lab{smallw}
\ee
By taking now a $U(1)$ subalgebra of \rf{ordca}
\be
\lcurl {\cal J}(x) \, ,\, {\cal J}(y) \rcurl = k g_{\scriptstyle {\cal J}}
\d^{\pr}(x-y) \ee
one sees that the $(T,{\cal J})$ system now generates an algebra which is
isomorphic to  \rf{2jbarj} for $h=0$ and $c=  k g_{\scriptstyle {\cal J}}$.
The quantities $w_n$ introduced above will then generate a $c$-deformed
(nonlinear) area preserving diffeomorphism algebra, which corresponds to
\rf{third} with $h=0$, i.e.
\br
\lcurl w_{n}(x) \, ,\, w_{m} (y) \rcurl &=&
\( n + m -2 \) w_{n+m-2} (x) \d^{\pr} (x - y)
+( m-1 ) \( w_{n+m-2 } (x)\)^{\pr} \d (x -y) \nonumber\\
&-& k g_{\scriptstyle {\cal J}} (n-2)(m-2)w_{n-1}(x) \pa_x ( w_{m-1}(x)
\d (x-y))     \lab{smallwc}
\er
Another realization of the algebraic structure discussed in section 2 is found
within the context of the two-loop Kac-Moody algebra introduced in \ct{toda}
and further studied in \ct{schwimmer}
\br
\lb J^m_a (x) \, , \, J^n_b (y) \rb &=& f_{ab}^c J_c^{m+n} \d
(x - y)  + k g_{ab} \pa_x  \d (x - y)  \delta_{m,-n}
+ J^C (x) \d (x - y)  g_{ab} m \delta_{m,-n} \phantom{......}
\nonu \\
\lb J^D (x) \, , \, J^m_a (y) \rb &=& m J^m_a (y) \d (x -y)
\nonu\\
\lb J^C (x) \, , \, J^D (y) \rb &=& k \pa_x  \d (x - y)
\nonu \\
\lb J^C ( x) \, , \, J^m_a (y) \rb &=& 0 \lab{kma}
\er
The currents $J^C$ and $J^D$ satisfy the same algebra as \rf{1jbarj}.
Therefore  the two-loop WZNW model possesses the algebraic structure
discussed in subsection \rf{firstkp}.
In addition its Sugawara tensor
together with suitably chosen spin one currents generate the symmetries
discussed above for the case of ordinary WZNW models. In particular when the
spin one current is $J^C$ the symmetries are those discussed in \ct{deform}.

Unlike the ordinary WZNW model its two-loop version possesses the algebra
\rf{2jbarj} with $h\neq 0$. This is realized as follows. The Sugawara tensor
can be modified, in ordinary or two-loop WZNW, by adding a derivative of a spin
one current in the Cartan subalgebra.
The effect of this is to change the conformal spin of the currents
and to add, for
some currents, a $\d^{\pr \pr}$ term on the r.h.s. of \rf{primary}
(producing therefore $h\neq 0$), and also to add an anomaly term
$\d^{\pr \pr \pr}$ on the r.h.s. of \rf{virsuga}.
This last anomaly is unwanted because it spoils the algebraic
structure we are interested in. In the case of the two-loop WZNW model such
anomaly can be avoided by choosing the current modifying the  Sugawara tensor
to
be orthogonal to itself under the Killing form.
Such current can either be $J^C$ or $J^D$.
For the ordinary WZNW associated to a simple Lie group
there is no such current since the
Killing form restricted to the Cartan subalgebra is an Euclidean metric.

Summarizing, by modifying the Sugawara tensor of the two-loop WZNW model
\be
T_{\rm 2loop} (x) = {1 \o 2 k} \Bigl( \sum_{a,b=1}^{{\rm dim}\, \lie}
\sum_{n=- \infty}^{\infty} g^{ab} J^n_{ a} (x) J^{-n}_{b} (x)  +
2 J^{D}(x) J^{C} (x) \Bigr) \lab{suga}
\ee
as
\be
{\cal L} (x) = T_{\rm 2loop} (x) + \a \pa_x j (x)
\lab{2loop}
\ee
where $j$ can either be $\jc$ or $\jd$ but not a linear combination of
them.
The relation \rf{primary} for the current $K (x) = \g_C \jd +
\g_D \jc$ will then be
\be
\lcurl {\cal L} (x) \, , \, K (y) \rcurl = K(x) \d^\pr (x-y)
+ h_j \d^{\pr \pr} (x-y)
\lab{prim2}
\ee
where $h_{\scriptstyle {\jc,\jd}} = k \a \gamma_{C,D}$.
The relation \rf{virsuga} will not be modified since $j$ is orthogonal
to itself, i.e.
\be
\lcurl {\cal L} (x) \, , \, {\cal L} (y) \rcurl = 2 {\cal L} (x) \d^\pr (x-y)
+ {\cal L}^{\pr} (x) \d (x-y)
\lab{vir2}
\ee
we also have
\be
\lcurl K (x) \, , \, K (y) \rcurl = k ( \gamma_C + \gamma_D ) \d^\pr (x-y)
\lab{kk}
\ee
Notice that ${\cal L}$ and $K$ satisfies the algebra \rf{2jbarj} with
$h= h_j$ and $c=k \( \gamma_C + \gamma_D\) $.

\subsection{Toda Theories and {\bf $\bw$} Algebra}

Let us now say some words about the symmetries one gets when WZNW models are
reduced. When the ordinary non compact WZNW is reduced to the Conformal Toda
models (CT) \ct{ora} the Sugawara tensor is modified in order to preserve the
conformal symmetry. This modification introduces an anomaly in the Virasoro
algebra which spoils the structures leading to the area  preserving
diffeomorphism algebra \rf{smallw} and its $c$-deformed version
\rf{smallwc}.
After reduction, the quantities $w_n= T {\cal J}^{n-2}$
given above, will probably be non local, and will
not satisfy a simple algebraic structure.
It would be interesting to explore what sort of structure the
area preserving symmetry of
the ordinary WZNW leads to in the CT models after reduction.

In the case of the reduction of the two-loop WZNW model to the Conformal
Affine Toda models (CAT) \ct{toda} some of the symmetries discussed above
survive. This is true mainly because the $J^C$ current is untouched by the
reduction procedure, and so is a local current for the CAT model.
In order to preserve the conformal symmetry
the Sugawara tensor \rf{suga}
is modified as \ct{toda}
\be
L (x) = T_{\rm 2loop} (x) + \pa_x \( 2 J_{\hat \delta} (x) +
h J^{D }(x) \)
\lab{modsuga}
\ee
where $J_{\hat \delta}=k \Tr \( {\hat g}^{-1} \pa_{+} {\hat g} \,{\hat
\delta}\cdot H^{0} \)$ with ${\hat \delta} = \h \sum_{\a>0} \a /\a^2$ and $h$
is the Coxeter number of the underlying semisimple Lie algebra.
Such tensor satisfies the Virasoro algebra with an anomaly
\be
- 4 \( {\hat \delta}\)^2 k \, \delta^{\pr\pr\pr} (x - y)
\lab{central}
\ee
An important point is that the $J^C$ current can be used to
modify further the tensor \rf{modsuga} in order to cancel such anomaly.
We then define
\be
U (x) \equiv L(x) + {2 {\hat \delta}^2 \o h} \, \pa_x \jc (x)
\lab{utensor}
\ee
which is still conserved and satisfies
\be
\{ U (x) \, , \, U (y) \} =2 U (x) \d^{\pr}
(x - y) +  U^{\pr} (x) \delta (x - y)
\lab{vira3}
\ee
In addition
\br
\{ U (x) \, , \, J^C (y) \} &=& J^C (x) \d^{\pr}
(x - y) + hk \d^{\pr\pr} (x - y)
\lab{wpi}\\
\{ J^C (x) \, , \, J^C (y) \} &=& 0 \lab{abel}
\er
The algebra \rf{vira3}-\rf{abel} gives rise to the $\bw$ algebra like
under the second bracket \rf{2jbarj}, with $c=0$.

Therefore the CAT
model not only possesses an area preserving symmetry but also
a richer structure as discussed in \ct{deform}.
Let us remark that the CAT model does not possess
an algebra of the type \rf{1jbarj}.
The currents $J^C$ and $J^D$ of the two-loop WZNW satisfy the algebra
\rf{1jbarj}. However after the reduction the current $J^D$ becomes non local
in the CAT model field variables. The algebra of $J^C$ and $J^D$ is then non
local and certainly not isomorphic to \rf{1jbarj}. Another point is that the
algebra \rf{1jbarj} does not possess highest weight unitary representations
\ct{schwimmer}. Therefore such symmetry would jeopardize
the hope of the CAT model being unitary at the quantum level.

\subsection{KP Hierarchy and One-Matrix Model}
Here we establish a connection between our realization of Section 3
with the construction of the KP hierarchy associated with
the Discrete Linear System:
\br
Q \Psi &=& \l \Psi                 \lab{qpsi}\\
\partder {\Psi}{t_r} &=& Q^r_+\Psi      \lab{trpsi}
\er
proposed recently by Bonora and Xiong in \ct{alternative}.
In \rf{qpsi} and \rf{trpsi}
\be
\Psi \equiv \( \begin{array}{c}
\vdots \\
\Psi_{n-1} \\
\Psi_{n} \\
\Psi_{n+1} \\
\vdots \end{array} \)
\qquad;\qquad
Q=I_{+}  +  \sum_{i=0}^{\infty}  a_i I_{-}^i
\lab{qii}
\ee
with the raising/lowering matrices $I_{\pm}$ from eq. \rf{iplus}.

For completness we summarize the major steps taken in \ct{alternative}.

The compatibility condition of equations \rf{qpsi} and \rf{trpsi} gives rise
to the discrete KP hierarchy i.e.,
\be
\partder {Q}{t_r}  =  \lb  Q^r_{+} \, , \, Q \rb
\lab{dkp}
\ee
Taking $r=1$ in \rf{trpsi} one gets (with $\pa \equiv \pa / \pa t_1$)
\be
\pa \Psi_n = \Psi_{n+1} +  a_0 (n) \Psi_n
\lab{psin}
\ee
which can be rewritten as
\be
\Psi_{n+1} = \( \pa - a_0 (n) \) \Psi_n \; \longrightarrow\;
\Psi_n = \hb_n \Psi_{n+1}\; \;\; , \;\;\; \hb_n \equiv
{ 1 \o \pa - a_0 (n) }
\lab{hbn}
\ee
In this notation we can describe the action of $Q$ on $\Psi$ as
\br
\( Q \Psi \)_n &=& \Psi_{n+1} + \sum_{i=0}^{\infty} a_i \hb_{n-i}
\hb_{n-i+1} \ldots \hb_{n-1} \Psi_n  \nonu\\
&=& \pa \Psi_{n} + \sum_{i=1}^{\infty} a_i \hb_{n-i}
\hb_{n-i+1} \ldots \hb_{n-1} \Psi_n
\lab{qonpsi}
\er
which allows rewriting $Q$ in the space of $\Psi$'s as a differential
operator $\hq$ labeled by $n$
\be
\hq_n \Psi_n =\( Q \Psi \)_n \;\; \longrightarrow \;\;
\hq_n \equiv \pa + \sum_{i=1}^{\infty} a_i \hb_{n-i}
\hb_{n-i+1} \ldots \hb_{n-1}
\lab{hqn}
\ee
Rewritting the Discrete KP hierarchy \rf{dkp} in terms of components
$a_i (n)$ we get for the first flow:
\be
\partder {a_i(n)}{t_1} = a_{i+1} (n+1) - a_{i+1} (n) + a_i (n)
\( a_0 (n) - a_0 (n-i) \)
\lab{1flowai}
\ee
The form of this equation is such that the choice
$a_i (n) = 0, \quad \forall i\geq 2$ at $t_1 =0$ will be preserved.
In addition, we denote
\be
a_0 (n) = S_n\quad\; , \quad\; a_1(n) = R_n
\lab{restrict}
\ee
The flow equation \rf{1flowai} gives therefore
\br
\partder {S_n} {t_1} & = & R_{n+1}-R_n  \lab{snt1}\\
\partder {R_n}{t_1} &=& R_n \(S_n - S_{n-1} \) \lab{rnt1}
\er
for $n \in \IZ$.
One recognizes in from \rf{qii} and \rf{restrict} the Jacobi matrix
relevant for one-matrix models \ct{bmx}.
In order to
connect with the one-dimensional Toda model
we insert $S_n = - \pa \varphi_n/ \pa t_1$ in the flow
equation \rf{rnt1}. This results in
\be
R_n = \exp \({\varphi_n - \varphi_{n-1} }\) \nonu
\ee
Now \rf{snt1} takes the form of one-dimensional Toda equation
\be
{\pa^2 \varphi_n \o \pa t_1^2} = \exp \({\varphi_n - \varphi_{n-1} }\)
- \exp \({\varphi_{n+1} - \varphi_{n} } \)
\lab{toda1}
\ee
The corresponding truncated $\hq_n$ operator becomes
\be
\hq_n = \pa +R_n \hb_{n-1} = \pa +R_n {1 \o \pa - S_{n-1}}
= \pa +R_n {1 \o \pa - S_{n}+ R^{\pr}_n / R_n}
\lab{trunhq}
\ee
where in the last identity use was made of the flow equation \rf{snt1} in
order to express $\hq_n$ in terms of the fields taken at the same site $n$.
We can rearrange $\hq_n$ in a way which we will allow us to make
a contact with our $(\bj,J)$ approach to KP hierarchy.
Consider
\be
\pa \( R_n {1 \o \pa - S_{n}+ R^{\pr}_n / R_n} \) = R_n^{\pr}
{1 \o \pa - S_{n}+ R^{\pr}_n / R_n} + R_n \pa {1 \o \pa - S_{n}+
R^{\pr}_n / R_n}
\nonu
\ee
adding and subtracting $\(- R_n S_{n} {1 \o \pa - S_{n}+ R^{\pr}_n / R_n}
\)$ to the right hand side of the last identity we arrive after some
algebra at
\be
R_n \hb_{n-1}\; =\; \hb_{n} R_n
\lab{rnhbn}
\ee
Hence the $\hq_n$ operator can be written as
\be
\hq_n = \pa +\hb_{n} R_n = \pa + {1 \o \pa - S_{n}} R_n
\lab{newhqn}
\ee
in which we recognize the Lax operator introduced in \rf{conjlax}
when we identify $S_n = J$ and $R_n = \bj$.

Another method to express $\hq_n$ by variables from the same $n$-sector
would be to use \rf{snt1} to write $\hq_n = \pa + ( R_{n-1} +
S^{\pr}_{n-1}) \hb_{n-1}$.
Here we make the identification $S_{n-1} =- J$ and $R_{n-1} = \bj$.
We obtain in this way the Lax operator $D + (\bj - J^{\pr})(D + J)^{-1}$
with coefficients of the form $(-1)^n (\bj - J^{\pr} ) P_n (J)$.
In fact the change $\bj \to \bj - J^{\pr}$ preserves the algebra
\rf{1jbarj}. Also the form of \rf{2jbarj} is preserved
under this transformation provided $c=-2h$ (but with $h \to -h$).

Hence with fields $R_k$ and $S_k$ in $\hq_n$ appearing in the same
$n$ or $n-1$ sector we obtain the Lax representation for KP hierarchy
identical to the systems associated with the first row or first column
of our matrix \rf{wmatrix}.

\appendix
\section{\faa Polynomials. Definitions and Identities}
\lab{appendixa}
\setcounter{equation}{0}

In 1855 \faa considered the following problem \ct{faa}.
Let variables $x$ and $y$ be related through a smooth function
$\p$: $x= \p (y)$. Let $f$ be another smooth function of $x$.
Find closed expression for the n-th derivative of $f$ with respect to
$y$ or in another words find how does $\pa^n f$ behave under change of
variables $x \to y$.
The formula which \faa found to provide an answer to the above problem
involves a nice piece of combinatorics and is given in terms of Bell
polynomials $B_{n,k}$ by:
\br
\pa_y^n f &=& \sum_{k=1}^n \pa_x^k f \; B_{n,k} \( \p^{(1)}, \p^{(2)},
\dots,\p^{(n-k+1)} \) \lab{def} \\
B_{n,k} \( x_{1}, x_{2},\dots,x_{l} \) &\equiv&
\sum { n ! \o i_1! i_2! \ldots i_l !}  \({x_1 \o 1!}\)^{i_1}
\({x_2 \o 2!}\)^{i_2} \ldots \({x_l \o l!}\)^{i_l} \lab{bell}
\er
where $\p^{(i)} = \pa^i \p$ and summation is over integers
$i_1,i_2, \ldots ,i_l\geq 0$ satisfying
\br
i_1 + 2 i_2 + 3 i_3 + \ldots+ l i_l &=& n \nonu \\
i_1 +  i_2 +  i_3 + \ldots + i_l &=& k \nonu
\er
where clearly $l_{\rm max} = n-k+1$.

In order to make connection to definition of \faa polynomials used in
this paper let us take the special function $f (x) = \exp (x)$ and
multiply both sides of \rf{def} by $f^{-1}$ with the result
\br
P_n (J) &\equiv& e^{-\p} \pa^n e^{\p} = \sum_{k=1}^n B_{n,k} (J)
\lab{faadef} \\
&=& \sum_{i_1 +2 i_2+ \ldots =n}
 { n ! \o i_1! i_2! \ldots i_l !}  \({J \o 1!}\)^{i_1}
\({J^{\pr} \o 2!}\)^{i_2} \ldots \({J^{(l-1)} \o l!}\)^{i_l}
\;\; ;\;\; J \equiv \p^{\pr} \nonu
\er
where $B_{n,k} (J) \equiv B_{n,k} \(J, J^{\pr}, J^{\pr\pr}, \ldots \)$.
As a corollary of \rf{faadef} we can find a
generating function for \faa polynomials.
Consider namely:
\be
e^{\p (x+\eps)- \p (x)}= \exp \lcurl \sum_{k=1}^{\infty} {\eps^k \o k!}
J^{(k-1)} \rcurl
\lab{ephi}
\ee
On the other hand applying Taylor expansion to $e^{\p (x+\eps)}$
we arrive at an expression for the generating function:
\be
\exp \lcurl \sum_{k=1}^{\infty} \eps^k J^{(k-1)}/k! \rcurl
 = \sum_{k=0}^{\infty} P_k (J) \eps^k /k!
\lab{schur}
\ee
{}From equation \rf{schur} one can easily prove an identity
\be
P_n \( f + g \) = \sum_{k=0}^n {n \choose k} P_k (f) P_{n-k} (g)
\qquad;\qquad n=0,1,\ldots
\lab{fgid}
\ee
valid for arbitrary functions $f$ and $g$.
An alternative expression for \faa polynomials is:
$P_n (J) = (D +J)^n \cdot 1 $ leading to a recurrence relation
$\pa P_n = P_{n+1} - J P_n$.
Few lowest order polynomials are listed here for convenience:
\be
P_0  = 1 \;\; ;\;\; P_1 = J \; \; ; \;\;
P_2 =  J^{2} +  J^{\pr}  \; \; ;\;\;
P_3 =    J^{3} + 3 J J^{\pr} +  J^{\pr \pr}  \;\;\; {\rm etc}.
\lab{lowest}
\ee

The exponential representation in \rf{faadef} turns out to be
useful in proving several identitities involving Fa\'{a} di Bruno
polynomials.
First we recall the following basic formulas:
\br
\pa^n f &=& \sum_{\a=0}^n (-1)^{\a} {n \choose \a} D^{n-\a} f D^{\a}
\lab{panf}\\
D^n f &=& \sum_{\a=0}^n {n \choose \a} \pa^{n-\a} f D^\a \lab{dnf}\\
f D^n &=& \sum_{\a=0}^n (-1)^{\a} {n \choose \a} D^{n-\a} \(\pa^\a f\)
\lab{fpan}\\
D^{-n} f &=& \sum_{\a=0}^{\infty} (-1)^{\a} {(n +\a-1)! \o \a ! (n-1)!}
\, \pa^{\a} f D^{-n-\a} \lab{dminusnf}
\er
where $\pa^k f \equiv f^{(k)}$ and $D={\pa /\pa x}$ acts to the right
as a derivative operator.

We derive now a number of useful identities (most of them are listed in
\ct{dickey}).
Using \rf{dnf} we find:
\be
\(D +J\)^n = e^{-\p} D^n e^{\p} = \sum_{l=0}^n {n \choose l}
e^{-\p} \pa^{n-l} e^{\p} D^l = \sum_{l=0}^n {n \choose l}
P_{n-l} (J) D^l \lab{751}
\ee
Using \rf{panf} we get:
\be
P_n (J) = e^{-\p} \pa^n e^{\p} = \sum_{l=0}^n (-1)^l {n \choose l}
e^{-\p} D^{n-l} e^{\p} D^l = \sum_{l=0}^n (-1)^l {n \choose l}
\(D +J\)^{n-l} D^l \lab{754}
\ee
while using \rf{fpan} we get:
\be
P_n (J) = e^{-\p} \pa^n e^{\p} = \sum_{l=0}^n (-1)^l {n \choose l}
D^{n-l} \( \pa^l e^{-\p}\) e^{\p} = \sum_{l=0}^n (-1)^l {n \choose l}
D^{n-l} \( (D - J)^l \dot 1 \) \lab{753}
\ee
{}From \rf{dnf} we also find:
\be
\(D-J\)^n = e^{\p} D^n e^{-\p} = \sum_{l=0}^n (-1)^{n-l} {n \choose l}
D^l \(\pa^{n-l} e^{\p}\) e^{-\p} = \sum_{l=0}^n (-1)^{n-l}
{n \choose l} D^l P_{n-l} (J) \lab{752}
\ee
Similarly we get:
\br
\lefteqn{
P_n(J) \( D-J \)^m =\pa^n e^{\p} D^m e^{-\p}}\nonu\\
&=&
\sum_{l=0}^m (-1)^{m-l} {m\choose l} D^l e^{-\p} \pa^{m+n- l} e^{\p}
= \sum_{l=0}^m (-1)^{m-l} {m\choose l} D^l P_{m+n- l} (J)
\lab{leonids}
\er
from which we get
\be
P_n(J)\sum_{k=0}^m (-1)^k {m\choose k} D^k P_{m-k}(J) =
\sum_{k=0}^m (-1)^k {m\choose k} D^k P_{n+ m-k} (J)
\lab{ourformula}
\ee
Finally let us also list for completeness
\br
D^n &=& D^n \( e^{\p} e^{-\p}\) =
\sum_{l=0}^n {n\choose l} e^{-\p} \( \pa^{n- l} e^{\p} \)
e^{\p} D^{l} e^{-\p} \nonu\\
&=& \sum_{l=0}^n {n\choose l} P_{l} (J) \( D - J \)^{n-l}
\lab{755}
\er
and
\br
D^n &=&\( e^{-\p} e^{\p}\) D^n =
\sum_{l=0}^n (-1)^{l} {n\choose l} e^{-\p} D^{n- l} e^{\p} e^{-\p}
\pa^l e^{\p} \nonu\\
&=& \sum_{l=0}^n (-1)^{l} {n\choose l} \( D +J \)^{n-l} P_{l} (J)
\lab{756}
\er
Note also from \rf{dminusnf} that
\be
e^{-\p}D^{-1} e^{\p} =
\sum_{l=0}^{\infty} (-1)^l e^{-\p} \pa^{l} e^{\p} D^{-l-1}=
\sum_{l=0}^{\infty} (-1)^l P_{l} (J) D^{-l-1}
\lab{dinverse}
\ee
and hence $ \(D+J\)^{-1} = e^{-\p}D^{-1} e^{\p} $.

\section{Appendix B }
\lab{appendixb}
\setcounter{equation}{0}
Here we give a proof of the identities \rf{id1}-\rf{id4}.
{}From \rf{u} we have
\br \lefteqn{
{\cal U}^{(r)}_{n+1,m}(x) + D_x {\cal U}^{(r)}_{n,m}(x) =  (-1)^r
\sum_{k=0}^{n+1} (-1)^k {n+1 \choose k} W^{(r)}_{n+m-k+1} D^k_x} \nonu\\
&+& (-1)^r \sum_{k=0}^{n} (-1)^k {n \choose k} (\pa_x W^{(r)}_{n+m-k}) D^k_x
+ (-1)^r \sum_{k=0}^{n} (-1)^k {n \choose k} W^{(r)}_{n+m-k} D^{k+1}_x
\nonu\\
&\equiv & I+II+III
\er
Making the shift $k+1\rightarrow k$ in $III$ and adding it to $I$ one gets
\be
(-1)^r (I+III) = W^{(r)}_{n+m+1} + \sum_{k=1}^{n+1} (-1)^k \biggl(
{n+1 \choose k} -  {n \choose k-1} \biggr) W^{(r)}_{n+m-k+1} D^k_x
\ee
The term $k=n+1$ does not contribute to the sum. Using then the identity
\be
{n+1 \choose k} = {n \choose k} + {n \choose k-1}
\lab{combid}
\ee
one notices that, as a consequence of \rf{k2kplusone}, the sum $I+II+III$ is
equal to ${\cal U}^{(r+1)}_{n,m}(x)$. So \rf{id1} is proven.

{}From \rf{v}
\br
 {\cal V}^{(r)}_{n+1,m}(x) + D_x {\cal V}^{(r)}_{n,m}(x) &=&
(-1)^r \sum_{k=0}^{m} {m \choose k} D^k_x W^{(r)}_{n+m-k+1}  \nonu\\
&+& (-1)^r \sum_{k=0}^{m} {m \choose k} D^{k+1}_x W^{(r)}_{n+m-k}
\equiv  I+II
\er
Make the shift $k+1\rightarrow k$ in $II$. Combine $I$, without the term $k=0$,
with $II$, without the term $k=m+1$, and use identity \rf{combid}. One then
gets that $I+II$ is equal to ${\cal V}^{(r)}_{n,m+1}(x)$. So, \rf{id2} is
proven.

The proof of \rf{id3} is very similar to that for \rf{id2}. Just notice
that $D_y$ will act directly on $\d (x-y)$ and then $D^k_x D_y \d
(x-y) = D^{k+1}_x \d (x-y)$.

{}From \rf{id2} we have
\be
{\cal V}^{(r)}_{n+1,m}(x) \d (x-y) + (\pa_x {\cal V}^{(r)}_{n,m}(x))
\d (x-y) - D_y {\cal V}^{(r)}_{n,m}(x)  \d (x-y) =
{\cal V}^{(r)}_{n,m+1}(x) \d (x-y)
\lab{almostthere}
\ee
Using \rf{k2kplusone} one has
\br
{\cal V}^{(r)}_{n+1,m}(x) \d (x-y) + (\pa_x {\cal V}^{(r)}_{n,m}(x))
\d (x-y) &= & (-1)^r \sum_{k=0}^{m} {m \choose k} D^k_x \biggl(
W^{(r)}_{n+m-k+1}(x)  \nonu\\
+\pa_x W^{(r)}_{n+m-k}(x) \biggr) \d (x-y)
&= &{\cal V}^{(r+1)}_{n,m}(x) \nonu
\er
Substituting this in \rf{almostthere} one gets \rf{id4}.

\section{Appendix C}
\lab{appendixc}
\setcounter{equation}{0}

We now prove the relation \rf{linealgebra}.
By applying the identity \rf{k2kplusone} successively one gets
\be
W^{(r)}_m (x) = (-1)^{m-n} \sum_{k=0}^{m-n} {m-n \choose k} \pa_x^{m-n-k}
W^{(r+k)}_n (x)
\lab{successive}
\ee
for $m \geq n$.
Then from \rf{u} one gets
\be
{\cal U}^{(r)}_{n+s,n} (x) = (-1)^{n+r+s} \sum_{k=0}^{n+s} \sum_{l=0}^{n+s-k}
{n+s \choose k} {n+s-k \choose l} ( \pa_x^{n+s-k-l} W^{(r+l)}_n (x)) D_x^k
\ee
Now using the identity
\be
{m \choose k} {m-k \choose l} = {m \choose l} {m-l \choose k}
\ee
and the fact that
\be
\sum_{i=0}^m \sum_{j=0}^{m-i} = \sum_{j=0}^m \sum_{i=0}^{m-j}
\lab{doublesum}
\ee
one gets
\br
{\cal U}^{(r)}_{n+s,n} (x) &=& (-1)^{n+r+s} \sum_{l=0}^{n+s} \sum_{k=0}^{n+s-l}
{n+s \choose l} {n+s-l \choose k} (\pa_x^{n+s-k-l} W^{(r+l)}_n (x) ) D_x^k
\nonu \\
&=& (-1)^{n+r+s} \sum_{l=0}^{n+s}{n+s \choose l} D_x^{l}
W^{(n+r+s-l)}_n(x)
\lab{upart}
\er
where we have used \rf{dnf} and made the shift $n+s-l \rightarrow l$.

{}From \rf{v} and \rf{successive} we have
\br
{\cal V}^{(s)}_{n,n+s} (x) &=&
(-1)^{n+r+s} \sum_{k=0}^{n+r} \sum_{l=0}^{n+r-k} (-1)^k {n+r \choose k} {n+r-k
\choose l} D_x^k ( \pa_x^{n+r-k-l} W^{(s+l)}_n )  \\
&=& (-1)^{n+r+s} \sum_{k=0}^{n+r} \sum_{l=0}^{n+r-k} \sum_{m=0}^k (-1)^k {n+r
\choose k} {n+r-k \choose l} {k \choose m}  ( \pa_x^{n+r-m-l} W^{(s+l)}_n )
D_x^m \nonu
\er
where in the last equality we used \rf{dnf}. Now using \rf{doublesum}
for the double sum in $k$ and $l$ and also \be \sum_{k=0}^{n+r-l}
\sum_{m=0}^{k}
= \sum_{m=0}^{n+r-l} \sum_{k=m}^{n+r-l} \ee one gets
\br \lefteqn{
{\cal V}^{(s)}_{n,n+s} (x) =
(-1)^{n+r+s} \sum_{l=0}^{n+r} \sum_{m=0}^{n+r-l} {(n+r)! \over {l! m!}}
( \pa_x^{n+r-m-l} W^{(s+l)}_n )}\nonu\\
&\times& D_x^m  \biggl( {1 \over {(n+r-l-m)!}}
\sum_{k=m}^{n+r-l}{{(-1)^k (n+r-l-m)!}\over {(k-m)! (n+r-l-k)!}} \biggr)
\er
Making the shift $k-m \rightarrow k$ in the sum in $k$ and using
\be
\sum_{i=0}^m (-1)^i {m \choose i} = \d_{m,0}
\ee
one gets
\be
{\cal V}^{(s)}_{n,n+s} (x) =
(-1)^{n+r+s} \sum_{l=0}^{n+r} (-1)^l {n+r \choose l} W^{(n+r+s-l)}_n  D_x^l
\lab{vpart}
\ee
where we have made the shift $n+r-l \rightarrow l$.

Using \rf{upart} and \rf{vpart} one can easily get \rf{linealgebra}
from \rf{general}.

\section{Algebra for $\cw_n$ Generators}
\lab{appendixd}
\setcounter{equation}{0}

Here we will be interested in calculating brackets using the generating
functions from \rf{cweps}.
The relevant bracket is given by:
\br
\lefteqn{
\lcurl \tj (x) J (x+ \eps) e^{\D_{\eps} \Phi (x)}\, , \, \tj (y)
J (y+ \eta) e^{\D_{\eta} \Phi (y)} \rcurl =
e^{\D_{\eta} \Phi (y)} e^{\D_{\eps}\Phi (x)} \times }\lab{bigform}\\
&\times& \Bigg\lb \tj (y) J (x+\eps) \d^{\pr} (x-y-\eta)
- \tj (x) J (y+\eta) \d^{\pr}_y (y-x-\eps) \nonu \\
&-&\tj (y) J(y+\eta) J (x+\eps) \d (x-y-\eta)
+\tj (x) J (y+\eta) J (x+\eps) \d (y-x-\eps) \nonu \\
&+&\tj (x) \tj (y) J (y+\eta) \d (x+\eps-y)
-\tj (x)\tj (y) J (x+\eps) \d (y+\eta-x)\nonu\\
& -& 2 \tj (x) \tj (y) J (x+\eps) J (y+ \eta)
 \times \biggl( \veps (x+\eps-y-\eta) - \veps (x+ \eps -y)
\times \nonu \\
& - & \veps (x-y-\eta) + \veps (x-y) \biggr)\Bigg\rb \nonu
\er
The last term of \rf{bigform} will provide the nonlinear part:
\br
\lefteqn{
-2 \tj (x) \tj (y) \sum_{n,m,l,p}
\bigl( e^{-\Phi (x)} \pa^{n-l} J (x) e^{\Phi (x)} { \eps^{n-l} \o (n-l)!}
\bigr)
\bigl( e^{-\Phi (y)} \pa^{m-p} J (y) e^{\Phi (y)} { \eta^{m-p} \o (m-p)!}
\bigr) \times } \nonu\\
& &\biggl( \pa^l_x \pa^p_y \veps (x-y) { \eps^{l} \o l!} {\eta^{p} \o p!}
-\pa^l_x \veps (x-y) { \eps^{l} \o l!}
-\pa^p_y \veps (x-y) { \eta^{p} \o p!} + \veps (x-y) \biggr) \nonu\\
&=&-2 \tj (x) \tj (y) \sum_{n,m} \sum_{l,p=1}^{n,m} {n \choose l}
{m \choose p} (-1)^p
\bigl( e^{-\Phi (x)} \pa^{n-l} J(x) e^{\Phi (x)} \bigr)\times
\phantom{aaaaaa}\nonu\\
& &\bigl( e^{-\Phi (y)} \pa^{m-p} J (y) e^{\Phi (y)} \bigr)
\pa^{l+p-1}_x \d (x-y) { \eps^{n} \o n!} { \eta^{m} \o m!} \lab{epseta3}
\er

After working with delta function on the right hand side of
\rf{bigform} one finds
\br
\lefteqn{
\lcurl \tj (x) J (x+ \eps) e^{\D_{\eps} \Phi (x)}\, , \, \tj (y)
J (y+ \eta) e^{\D_{\eta} \Phi (y)} \rcurl = \phantom{aaaaaaaaaaaa}
}\lab{nbigform}\\
& & \tj (y) J (y+\eta +\eps) e^{\D_{\eta+\eps} \Phi (y)} \d^{\pr} (x-y-\eta)
+ \tj (x) J (x+\eta +\eps) e^{\D_{\eta+\eps} \Phi (x)} \d^{\pr} (x+\eps-y)
\nonu \\
&+& \d (x+\eps-y) \partder {} {(\eps + \eta)} \tj (x) J (x+\eta +\eps)
e^{\D_{\eta+\eps} \Phi (x)} \nonu\\
&-& \d(x-y-\eta) \partder {} {(\eps + \eta)} \tj (y) J (y+\eta +\eps)
e^{\D_{\eta+\eps} \Phi (y)} + {\rm last~term} \nonu
\er
by the last term we mean the last term of equation \rf{bigform},
which as shown in \rf{epseta3} after comparing with \rf{pnpmnonlin}
is reproducing the DS structure so for the time
being we do not need to worry about it.

We concentrate first on first and fourth terms on the r.h.s. of \rf{nbigform}.
Expanding in $\eps$ and $\eta$ gives for these two terms:
\be
 \sum_{n=0}^{\infty} \sum_{m=0}^{\infty} \sum_{s=0}^{m+1} {\eps^n \o n!}
{\eta^m \o m!} (-1)^{m+n} {m+1 \choose s} \cw_{m+n-s -1} (y) \d^{(s)} (x-y)
\lab{14}
\ee
Similarly for second and third terms on the r.h.s. of \rf{nbigform}
we find
\be
\sum_{n=0}^{\infty} \sum_{m=0}^{\infty} \sum_{s=0}^{n+1} {\eps^n \o n!}
{\eta^m \o m!} (-1)^{m+n-s-1} {n+1 \choose s} \cw_{m+n-s -1} (x)
\d^{(s)} (x-y)
\lab{23}
\ee
Adding to this the last term we finally obtain the following bracket
\br
\lefteqn{
\lcurl \cw_{n}(x) \, ,\, \cw_{m} (y) \rcurl =
\sum_{s=0}^{m+1} {m+1 \choose s} D^s_x \cw_{m+n-s +1} (x)
\d (x-y) }\lab{wuyud}\\
&-& \sum_{s=0}^{n+1} (-1)^s {n+1 \choose s} \cw_{m+n-s -1} (x) D^s_x
\d (x-y) \nonu\\
&-& 2 \Biggl( \sum_{i=0}^{n-1} \sum_{j=0}^{m-1} (-1)^{n-i} {n \choose i}
{m \choose j} \cw_{i}(x) D^{n+m-i-j-1}_x \cw_{j} (x)   \Biggr) \d (x-y)
\nonu
\er
where we copied the results for the last term from previous calculations.
The first two terms reproduce $\O^{(1)}_{nm} (\cw (x))$.

\lskip
{\bf Acknowledgements:}\\
We would like to thank Prof. L. Bonora for interesting discussions on
his work.
One of us (HA) thanks IFT-UNESP for hospitality.
We also acknowledge financial support within CNPq/NSF
Cooperative Science Program.
\lskip

\end{document}